\documentclass[%
reprint,
superscriptaddress,
showpacs,preprintnumbers,
amsmath,amssymb,
pra,
notitlepage,
]{revtex4-1}

\usepackage[toc,page]{appendix}
\usepackage{array}
\usepackage{graphicx}   
\graphicspath{{fig/}}   
\usepackage{dcolumn}
\usepackage{bm}
\usepackage{hyperref}

\usepackage{epstopdf}
\usepackage{braket}
\usepackage{mhchem}
\usepackage{siunitx}

\usepackage{changes}

\begin{document}

\title{Probing the Tavis-Cummings level splitting with intermediate-scale superconducting circuits}

\author{Ping Yang}
\affiliation{Institute of Physics, Karlsruhe Institute of Technology, 76131 Karlsruhe, Germany}
\author{Jan David Brehm}
\affiliation{Institute of Physics, Karlsruhe Institute of Technology, 76131 Karlsruhe, Germany}
\author{Juha Lepp\"{a}kangas}
\affiliation{Institute of Physics, Karlsruhe Institute of Technology, 76131 Karlsruhe, Germany}
\affiliation{HQS Quantum Simulations GmbH, 76131 Karlsruhe, Germany}
\author{Lingzhen Guo}
\affiliation{Max Planck Institute for the Science of Light,  91058 Erlangen, Germany}
\author{Michael Marthaler}
\affiliation{HQS Quantum Simulations GmbH, 76131 Karlsruhe, Germany}
\affiliation{Institute for Theoretical Condensed Matter physics, Karlsruhe Institute of Technology, 76131 Karlsruhe, Germany}
\affiliation{ Theoretische Physik, Universit\"at des Saarlandes, 66123 Saarbr\"ucken, Germany}
\author{Isabella Boventer}
\affiliation{Institute of Physics, Karlsruhe Institute of Technology, 76131 Karlsruhe, Germany}
\affiliation{Institute of Physics, University Mainz, 55128 Mainz, Germany}
\author{Alexander Stehli}
\affiliation{Institute of Physics, Karlsruhe Institute of Technology, 76131 Karlsruhe, Germany}
\author{Tim Wolz}
\affiliation{Institute of Physics, Karlsruhe Institute of Technology, 76131 Karlsruhe, Germany}
\author{Alexey V. Ustinov}
\affiliation{Institute of Physics, Karlsruhe Institute of Technology, 76131 Karlsruhe, Germany}
\affiliation{Russian Quantum Center, National University of Science and Technology MISIS, Moscow 119049, Russia}
\author{Martin Weides}
\email{martin.weides@glasgow.ac.uk}
\affiliation{Institute of Physics, Karlsruhe Institute of Technology, 76131 Karlsruhe, Germany}
\affiliation{James Watt School of Engineering, University of Glasgow, Glasgow G12 8QQ, United Kingdom}


\date{\today}


\begin{abstract}
We demonstrate the local control of up to eight two-level systems interacting strongly with a microwave cavity.
Following calibration, the frequency of each individual two-level system (qubit) is tunable without influencing the others.
Bringing the qubits one by one on resonance with the cavity, we observe the collective coupling strength of the qubit ensemble.
The splitting scales up with the square root of the number of the qubits, which is the hallmark of the Tavis-Cummings model.
The local control circuitry causes a bypass shunting the resonator, and a Fano interference in the microwave readout, whose contribution can be calibrated away to recover the pure cavity spectrum. The simulator's attainable size of dressed states with up to five qubits is limited by reduced signal visibility, and -if uncalibrated- by off-resonance shifts of sub-components.  Our work demonstrates control and readout of quantum coherent mesoscopic multi-qubit system of intermediate scale under conditions of noise.
\end{abstract}


\maketitle

\section{\label{sec:introduction}{Introduction}}
Most of today's quantum information systems rely on an interplay between an artificial atom and a resonator mode used for readout \cite{blais2004cavity}.
In absence of dissipation, its dynamics is well described by the Jaynes-Cummings model \cite{jaynes1963comparison}. For $N$ atoms interacting with one resonator Tavis and Cummings predicted a $\sqrt{N}$ enhancement of the effective coupling strength at degeneracy, leading to a level repulsion with a frequency gap $2g\sqrt{N}$, where $g$ is the coupling strength of one artificial atom to the resonator \cite{tavis1968exact}. After early experimental realizations with trapped ions~\cite{bernardot1992vacuum}, the $\sqrt{N}$-enhancement has been demonstrated with three locally tunable superconducting transmon qubits~\cite{fink2009dressed}, followed by eight qubits in a globally controlled ensemble ~\cite{macha2014implementation}, and a comparable number of transmons ~\cite{shulga2017observation}.\\
Novel applications have been proposed involving more than one controllable two-level system coupled to a single resonator.
These include bus systems realizing a tunable long-range interaction between distant qubits ~\cite{fillip2011qqcoupling,Majer2007qqcoupling,Sillanp2007qmemomy} and a quantum von Neumann architecture~\cite{Mariantoni61}. The collective interaction also creates multi-qubit entanglement~\cite{solano2007qubitentanglement} and provides protection against radiation decay~\cite{fillipp2011subradiance}. This versatility supports the use of Tavis-Cummings systems in future quantum simulators and computers.
An analog quantum simulation~\cite{Feynman1982analoguesimulation,georgescu2014qsim} of a Dicke model~\cite{Dicke1954} (generalized Tavis-Cummmings model) would provide a direct access to eigenenergies and transient dynamics of light-matter interaction in the ultra-strong coupling regime~\cite{Braumller2017,Anton2018}.\\
Experimentally, the increase of circuit complexity is a growing challenge for system control, e.g., due to cross-talk or circuit topology.  Global control of a large number of qubits is adversely affected by disorder in the ensemble, such as local flux offsets, which can be mitigated by local controls. Ideally, a residual finite interaction between the qubits themselves and cross-talk between flux lines can be calibrated away.\\
In this work, we increase the circuit complexity to study the Tavis-Cummings circuit consisting of a superconducting microwave resonator interacting with up to eight individually frequency-controllable transmon qubits. It is a well suited platform to study both desired and parasitic effects occurring in scaled-up quantum circuits.
We show a calibration method allowing for local qubit control of all eight qubits, and demonstrate its adequate analog quantum simulation of the Tavis-Cummings system by measuring the $\sqrt{N}$ coupling enhancement as the hallmark signature. Our circuit complexity is positioned between well-understood few qubit-resonator systems and their scaled up versions constructed to achieve quantum advantage. The experiment contains key properties such as decoherence, local control and crosstalk, reactive and dissipative background, and higher qubit levels, all of which are subtle features of any near-term physical quantum simulator.
\section{\label{sec:sample}{Multi-qubit chip and setup}}

\begin{figure} 
\includegraphics[width=1.0\columnwidth]{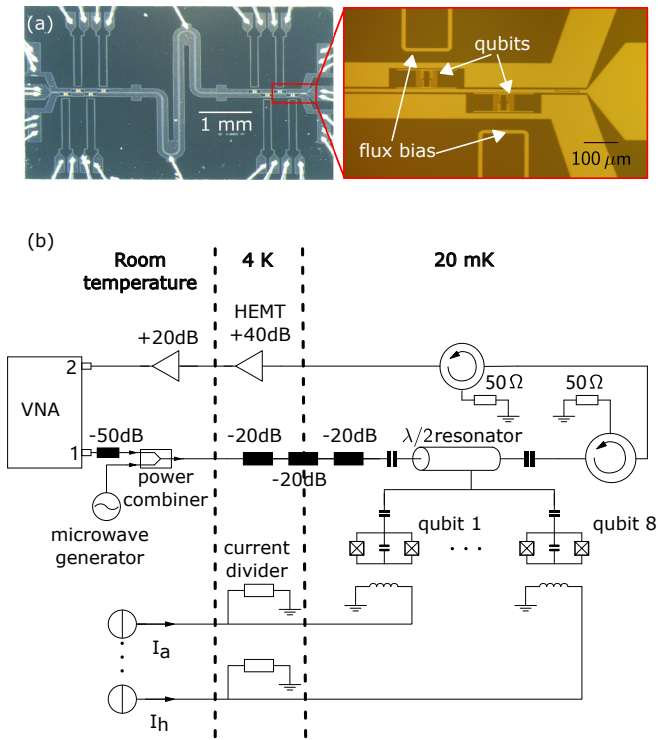}
\caption{(a) Optical micrograph of the chip bonded to the sample box. The meander-structure coplanar resonator is coupled at each end to four transmon qubits. U-shaped leads carry DC current to control the local magnetic flux. The enlarged image (red rectangle) shows two cavity-embedded transmon qubits and their flux bias lines. (b) Schematic of the measurement setup. The tone generated by the VNA is attenuated at different temperature stages of the refrigerator to reach single-photon regime and to lower the thermal noise at the quantum chip.}\label{FIG1measurementSetup}
\end{figure}

The quantum chip studied in this work contains a coplanar waveguide half-wavelength resonator with four transmon qubits capacitively coupled to each end of the resonator, as shown in Fig.~\ref{FIG1measurementSetup} (a), for a maximal coupling strength to each qubit. Each qubit frequency is individually controlled by a local DC flux bias. The theoretical description is given by the Tavis-Cummings model~\cite{tavis1968exact} with Hamiltonian
\begin{equation}\label{T-C}
 \hat{H}/\hbar=\omega_{\rm r}\hat{a}^{\dag}\hat{a}+\displaystyle{\sum_{i}}\frac{\omega_{i}}{2}\hat{\sigma}_{z_{i}}+\displaystyle{\sum_{i}}g_{i}(\hat{a}^{\dag}\hat{\sigma}_{i}^{-}+\hat{a}\hat{\sigma}_{i}^{+}).
\end{equation}
Here $\omega_{\rm r}$ is the pure resonator frequency, $\omega_i$ the qubit $i$ frequency, $g_i$ their coupling strength, and $\hat{\sigma}^{+,-}$ are the spin raising and lowering operators. The two-level systems are realized by transmon qubits~\cite{koch2007charge}, each including two Josephson junctions in a SQUID geometry enabling the local tunability by the applied magnetic fluxes.\\
Our experimental setup is illustrated in Fig.~\ref{FIG1measurementSetup} (b), further details, including the sample are given in the Appendix.\\
The microwave drive tone of a vector network analyzer (VNA) is attenuated along the signal chain by $120\,$dB in total. The power reaching the chip is $-137\,$dBm, where the average photon number on resonance is estimated to be $\braket{n}\approx 0.2$.\\
Ideally, the resonance frequency of the photon-dressed cavity can be observed as a peak in the microwave transmission spectrum.
In our experiments, the transmission data is characterized by asymmetric line-shapes, which implies an interference effect between the cavity and a background transmission, i.e. Fano resonance. The coupling to background modes is understood to emerge from a crosstalk with the multiple local control lines,
since it is absent in a single-qubit chip fabricated in the same run.
This Fano interference effect leads to an inverted spectrum, but does not affect the energy-level spacing~\cite{lepp2018Fano}.
Furthermore, once the background transmission is characterized from an off-resonance transmission,
its contribution on resonance can be subtracted from the transmission amplitude, resulting in the cavity spectrum only, see Appendix.\\
In Fig.~\ref{FIG2Calibration} (a) we show the measured power scan of microwave transmission across the cavity
when the Fano resonance is removed from the data. At low powers we observe the photon-dressed cavity frequency as a transmission peak.
With increasing power, the system enters a non-linear regime with the resonance frequency finally shifting towards
the bare cavity frequency. A local minimum appears at moderate powers, which has also been observed in Ref.~\cite{Reed2010}. It occurs due to an interference effect between two metastable states of the cavity at moderate drive powers~\cite{Bishop2010}.
In Fig.~\ref{FIG2Calibration} (b), we show data after removing the Fano-resonance at low transmission power when scanning the bias current to tune one transmon across the cavity resonance frequency.

\begin{figure} 
\includegraphics[width=1.0\columnwidth]{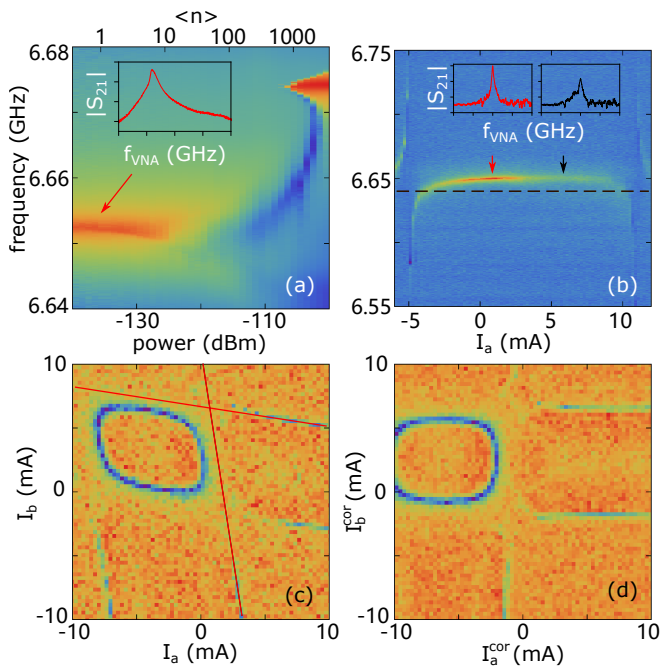}
\caption{(a) Power scan with all qubits far detuned (log-scale). The photon-dressed resonator frequency changes from low to high powers in transmission height and frequency. (b) Uncalibrated single-flux scan after background removal. Only one qubit is tuned through the cavity resonance frequency, all the other qubits are far detuned. The resonator frequency before background subtraction shows weakening of signal and transformation between peaks and dips~\cite{lepp2018Fano}. The horizontal black dashed line indicates the frequency point chosen to do the flux calibration. (c) Example for uncalibrated two-coil sweeps. The red lines are the fitted slopes which give the ratio between two mutual inductances. (d) Repeated measurement after calibration. The absence of a tilt indicates good isolation between the pair of flux lines.}\label{FIG2Calibration}
\end{figure}

\section{\label{sec:calibration}{Taking local control}}   
Being designed with local flux control of each qubit, the cross-talk between qubits and non-corresponding flux control lines is small, but not negligible, due to residual on-chip coupling, parasitic coupling in the DC wiring or within the DC current sources. For multi-qubit circuits with flux-tunable components careful calibration of the linear cross-talk is part of the experiment \cite{Harris2010, Weber2017}. This ensures true single-qubit control with DC current compensation routines on all the flux control lines.

Using one global readout, the calibration is fast, reproducible and -to some extent- scalable. Single-tone measurement of the resonator without exact qubit identification is sufficient to build the $8\times8$ mutual inductance matrix between flux control lines and qubits. The change of magnetic flux $\Delta \Phi$ through each qubit is calculated by:
\begin{equation}\label{mutual inductance}
  \begin{pmatrix}
    \Delta\Phi_{1} \\
     \vdots\\
    \Delta\Phi_{8} \\
  \end{pmatrix}=
  \begin{pmatrix}
   M_{1a} & M_{1b} && \cdots  & M_{1h} \\
   \vdots & \vdots && \ddots  & \vdots \\
   M_{8a} & M_{8b} && \cdots  & M_{8h} \\
  \end{pmatrix}
  \begin{pmatrix}
    \Delta I_{a} \\
         \vdots\\
    \Delta I_{h} \\
  \end{pmatrix},
\end{equation}
where $1, \ldots, 8$ label the qubits, and $a,\ldots, h$ indicate the flux bias. For instance, $\Delta\Phi_{1}$ is the flux variation through the first qubit, $\Delta I_{b}$ is the change of the DC current running through the second bias line, and $M_{1h}$ is the mutual inductance between the first qubit and the eighth flux bias line. Changing the current in one flux line does not only tune the frequency of its adjacent qubit, but also may bias other qubits. For calibration, a frequency close to the anti-crossing [see  black dashed line in Fig.~\ref{FIG2Calibration} (b)] is chosen. By observing the change in transmission while sweeping two bias currents, the mutual inductance matrix element is obtained. This value is used for the compensation currents to counteract the induced bias fluxes to effectively keep all other qubits at their frequencies. The corrected current after calibration $I_{i}^{\rm cor}$ is employed, rather than the absolute value of current $I_i$. Almost no-tilt indicates a good flux calibration, as seen in Fig.~\ref{FIG2Calibration} (d). For more details see Appendix.

\begin{figure} 
\includegraphics[width=1.0\columnwidth]{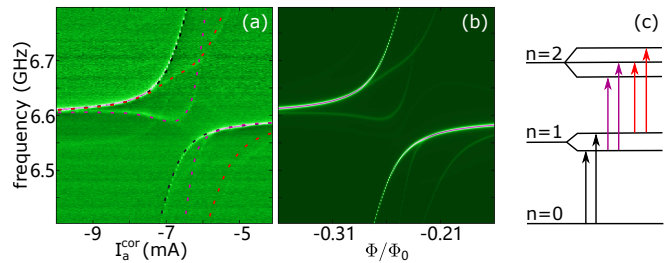}
\caption{Comparison between measurement and simulation of one transmon tuned through the cavity frequency. The plotted transmission amplitude is in log-scale. (a) Measured data of the anti-crossing.  The black lines correspond to excitations from the ground state and cause the vacuum Rabi splitting. The other colored lines correspond to higher-level transitions and are identified in (c). (b) Master equation simulation by QuTiP~\cite{johansson2012qutip} for a three-level artificial atom interacting with a resonator which has an average thermal photon population of 0.1 photons. (c) Energy-diagram of the first two excitation manifolds (schematic, not to scale) of the dressed system. }\label{FIG3higherLevel}
\end{figure}

\begin{table*}[tp!hbp]
\begin{tabular}{p{65pt}<{\hfil}|p{45pt}<{\hfil}|p{45pt}<{\hfil}p{45pt}<{\hfil}p{45pt}<{\hfil}p{45pt}<{\hfil}p{45pt}<{\hfil}p{45pt}<{\hfil}p{45pt}<{\hfil}p{45pt}<{\hfil}}
  \hline
  \hline
 & designed & qubit 1 & qubit 2 & qubit 3 & qubit 4  & qubit 5 & qubit 6 & qubit 7 & qubit 8 \\
  \hline
\scriptsize{$\omega^{\rm max}/2\pi\,$(GHz)} &  $9.11$  &  $7.90\pm0.05$   & $7.54\pm0.05$ & $7.70\pm0.04$  & $11.30\pm0.1$  & $10.10\pm 0.1$  & $9.70\pm0.08$  & $10.24\pm0.08$  & $12.22\pm0.08$  \\
\scriptsize{$g/2\pi\,$(MHz)} &  $113.0$  &  $114.8\pm0.2$   & $114.3\pm0.4$ & $113.4\pm0.6$  & $124\pm4$  & $107\pm 1$  & $110\pm1$  & $114.4\pm0.6$  & $109\pm4$  \\
  \hline
  \hline
\end{tabular}
\caption{Coupling strengths  $g_i$ and maximal energy-level splittings $\omega_{i}^{\rm max}$ of each transmon qubit.
Qubits 1-6 were used in the experiments probing the Tavis-Cummings level splitting.
$T_1$ times were estimated to vary between 50-80~ns and dephasing was limited by qubit decay, $T_2\approx T_1/2$~\cite{lepp2018Fano}.
The resonator coupling to transmission line is estimated to $\gamma\approx 2\pi\times 0.7$~MHz.
} 
\label{Tabel==coupling strength}
\end{table*}

\section{\label{sec:1qubit}{Individual qubit spectroscopy}}

Before probing the full Tavis-Cummings model, we determine the coupling strengths $g_i$ of each qubit from the minimal level-splitting, while parking all other qubits at their maximum frequencies. 
This level splitting is effectively described by eigenenergies $E_{\pm}/\hbar=\frac{\omega_{i}+\omega_{\rm r}}{2}\pm\frac{\sqrt{\Delta^{2}+4g_i^{2}}}{2}$, with
qubit frequency $\omega_i$ and $\Delta=\omega_{i}-\omega_{\rm r}$. 
At degeneracy ($\Delta=0$), the frequency difference $E_{\rm R}/\hbar$ is given by the vacuum Rabi splitting $\left(E_{+}-E_{-}\right)/\hbar=2 g_i$, i.e. the minimum distance between the major splittings (black dashed lines) as shown in Fig.~\ref{FIG3higherLevel} (a). The measured coupling strengths $g_i$ , see Table~\ref{Tabel==coupling strength}, indicate a good agreement between the designed and observed values.

The major splitting on resonance between one qubit and one resonator is described well by the Jaynes-Cummings model.
For detailed understanding of all features away from the Rabi splitting, the transmon has to be considered as a multi-level anharmonic oscillator, with $|{\rm g}\rangle,\ |{\rm e}\rangle,\ |{\rm f}\rangle$ denoting the first three uncoupled eigenstates respectively, and a Hamiltonian:
\begin{equation}\label{J-C under transmon base}
\hat{H}_{3{\rm L}}/\hbar=\omega_{\rm r}\hat{a}^{\dag}\hat{a}+\sum_{\substack{j=\\\rm g,e,f}}\omega_{j}|j\rangle\langle j|+\left(\hat{a}^{\dag}+\hat{a}\right) \sum_{\substack{i,j=\\\rm g,e,f}}g_{ij}|i\rangle\langle j|,
\end{equation}
in the base of $\{|{\rm g},0\rangle,\ |{\rm e},0\rangle,\ |{\rm f},0\rangle,\ |{\rm g},1\rangle,\ \cdots,\ |{\rm f},n\rangle\}$, and eigenenergies $\omega_{\rm g}$, $\omega_{\rm e}$ and $\omega_{\rm f}$. Only single-photon transitions between adjacent levels of the first three levels of transmon are taken into consideration, since two-photon transitions require much higher drive powers~\cite{braumuller2015multiphoton}. The analysis based on the (two-level) Jaynes-Cummings model, considered before, included transitions indicated by the two black arrows in Fig.~\ref{FIG3higherLevel} (c).
Including the third transmon state, additional transitions appear between the higher manifolds. All these six transitions are visualized in Fig.~\ref{FIG3higherLevel} (c) and plotted together with the measured anti-crossing in  Fig.~\ref{FIG3higherLevel} (a), next to the numerical  master equation simulation using QuTiP~\cite{johansson2012qutip} shown in Fig.~\ref{FIG3higherLevel} (b). We obtain a good agreement between the measured data and the model. This demonstrates detailed understanding of resonances appearing in our spectroscopy, and confirms that the additional features do not correspond to two-photon transitions (requring higher drive powers), but to single-photon transitions starting from the first, thermally excited, manifold. In combination with the low anharmonicity they can also cause additional vacuum Rabi splittings.

\section{\label{sec:6qubit}{Multi-qubit spectroscopy and $\sqrt{N}$ scaling of the coupling}} 

By bringing the transmons one by one on resonance with the cavity, we demonstrate the local control of multiple qubits and are able to measure the collective coupling.
The theoretical vacuum Rabi splitting generalizes to $E_{{\rm R}_{N}}/\hbar=\sqrt{\Delta^{2}+4Ng^{2}}$, assuming identical couplings $g_i=g$. When the $N$ qubits are exactly on resonance (i.e. $\Delta=0$), the splitting is $2g\sqrt{N}$~\cite{fink2009dressed}. Already for one qubit being slightly detuned, the measured splitting increases. Furthermore, considering $g_i$ being different for each qubit (even though relatively small), the Rabi splitting is given by $ E_{{\rm R_{N}}}/\hbar=2\sqrt{\sum_{i} g_{i}^{2}}$ at resonance.

\begin{figure}
\includegraphics[width=1.0\columnwidth]{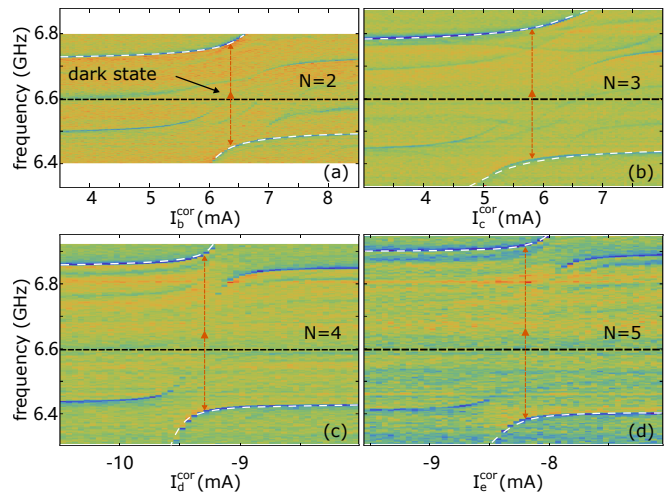}
\caption{ Multiple qubits on resonance with the cavity. Transmission amplitude in the log-scale for two (a) to five (d) qubits interacting resonantly with the cavity. The white dashed lines are fitting curves used to extract the collective coupling, see Appendix. The red dashed lines indicate the splittings, with the red triangle marking their centres (namely the resonator frequency). The black dashed lines show the resonator frequency when tuning only qubit 1 in resonance. }\label{FIG4qubitcoupling}
\end{figure}

\begin{figure}
\includegraphics[width=1.0\columnwidth]{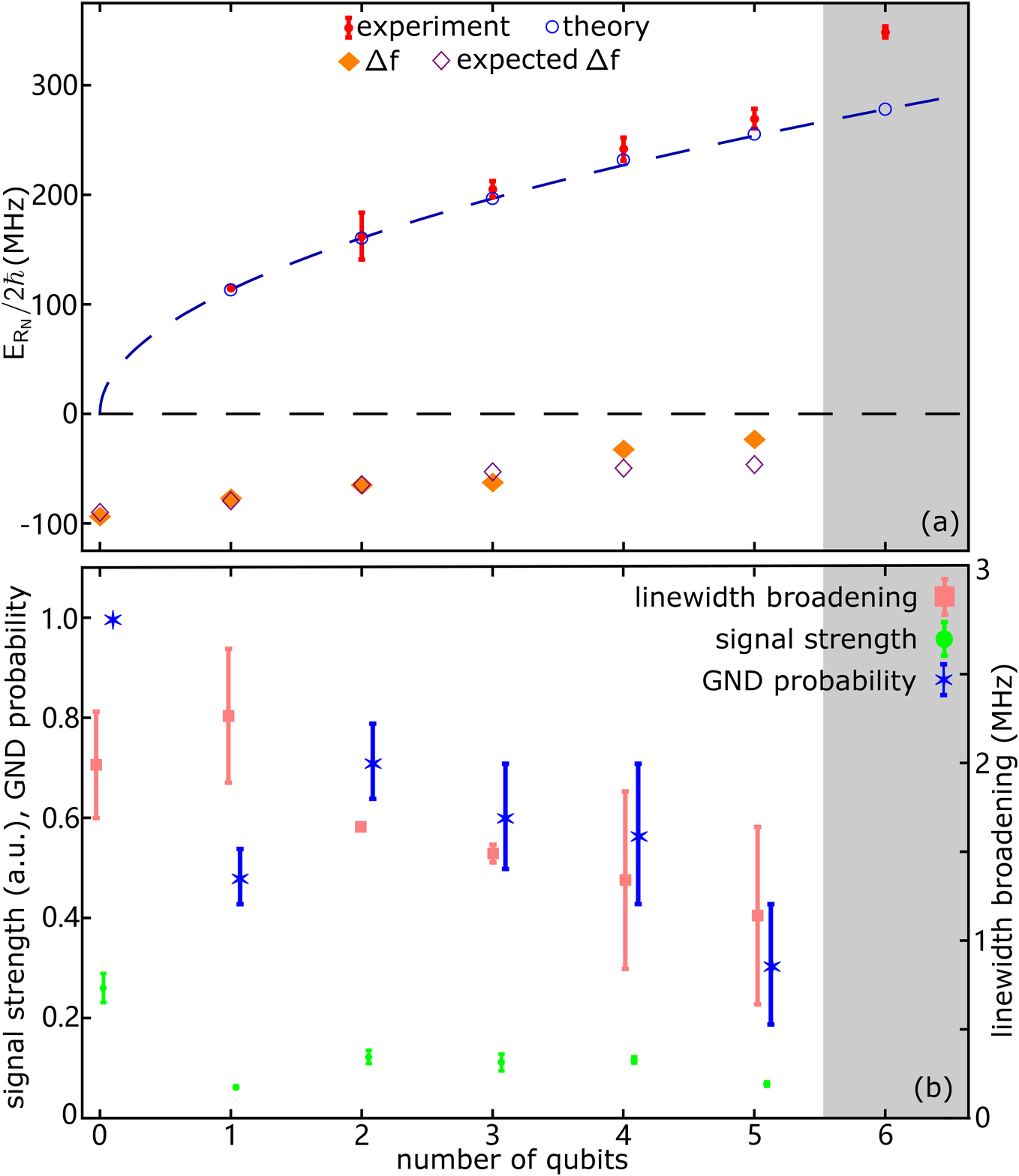}
\caption{Multiple qubits on resonance with the cavity.
(a) Comparison between theoretical (average measured individual coupling $g_{\rm avg}=\sum_i g_i/N$) and experimental vacuum Rabi splittings for $N$ qubits.
Up to $N=5$ calibration for all qubits has been applied, the $N=6$ data (grey area) is from uncalibrated measurement (red points).
Using the same reference, the rhombuses show the measured and expected shift of center frequencies $\Delta f$. 
The pure resonator frequency is calibrated to $0 \,\rm{MHz}$.
(b) Linewidth broadening $\kappa$, probability $p$ that the system is in the ground excitation-manifold, and the signal strength $|S_{21}|$ as function of total number of on-resonance qubits~$N$.}\label{FIG5qubitcoupling}
\end{figure}

Fig.~\ref{FIG4qubitcoupling} shows the transmission spectra revealing the collective vacuum Rabi splitting.
As a prerequisite, one qubit has been tuned into resonance, leading to an avoided level crossing as in Fig.~\ref{FIG3higherLevel} (a). In Fig.~\ref{FIG4qubitcoupling} (a), this qubit is kept on resonance, while the second qubit is tuned in.
$N=2$ qubits (and one resonator mode) have $N+1=3$ single-excitation eigenstates. One of the states is dark (no photons in the resonator), the excitation being shared only between the qubits. Similarly, we bring more qubits on resonance in (b) to (d). On resonance a bright doublet appears, corresponding to collective qubits-photon superposition states, $|N \pm\rangle=\frac{1}{\sqrt{2}}|{\rm g,..,g},1\rangle \pm \frac{1}{\sqrt{2N}}(|{\rm e,..,g},0 \rangle+...+|{\rm g,..,e},0 \rangle)$.
These eigenfrequencies are separated by the collective coupling $2g\sqrt{N}$.
The other $N-1$ single-photon excitation eigenstates are dark.

The measured bright doublets are fitted (white dashes in (a-d)) to extract the collective coupling strength and corresponding error.
For detailed information on the fitting procedure see Appendix.
A comparison to theoretical values indicated by measured individual couplings is shown in Fig.~\ref{FIG4qubitcoupling} (e).
All eight qubits are fully functional and tunable, although we manage to bring a maximum of six qubits on, or close, to resonance. 
For up to five qubits a good agreement to the prediction of the Tavis Cummings model is obtained.
A drift of the ensemble frequency, $\Delta f$, appears due to effective dispersive shifts $\sum_{i=N+1}^8 g_i^2/\Delta_i$ from off-resonance qubits.
The induced drift relative to the measured splitting is however minor as $|\Delta f/2g\sqrt{N}| \ll 1$.
The effective splitting and resonator drift $\Delta f$ are close the theoretical expectations, as shown in Fig.~\ref{FIG4qubitcoupling} (e).
The tuning precision is mainly limited by the steep flux dependence at the resonator frequency, in particular for high frequency qubits. For instance, a change of flux $\Delta\Phi=0.45\% \Phi_{0}$ (corresponding to $100.4\,\mu A$ in bias current) on qubit 5 results in a shift of $130\,$MHz in qubit frequency. The large flux susceptibility renders them sensitive to fluctuations in either the bias current (from the current source or picked up in the wire chain towards the qubits) or the magnetic background field. The qubit's larger linewidth reduces their signal strength, and therefore limits the tuning precision.

A central limiting factor in our experiment is the signal strength measured on-resonance in both traces, $|S_{21}| \propto p_N \gamma_N/(\kappa_N+\gamma_N)$.
We define it here by coupling to the transmission line $\gamma$, the linewidth broadening $\kappa$, and the probability for being in the ground excitation-manifold~$p$
(i.e., not being thermally excited out of the probed manifold).
These values depend on the bias points, here labelled simply by the amount of qubits brought on resonance~$N$.
The off-resonance coupling is $\gamma_0\approx 2\pi\times 0.7$~MHz and
for $N>0$ it halves to $\gamma_{N>0}=\gamma_0/2$ (due to the qubit-cavity hybridization, as a shared photon is only with 50 percent probability in the cavity).
The extracted signal strength and other parameters are plotted in Fig.~\ref{FIG5qubitcoupling}(b).
The values were determined for both the lower and the higher transmission peak at a resonance, whose difference defines the variance.

We observe that the signal strength drops strongly at $N=1$.
Here the degrading effect of elevated temperature and fast qubit decay becomes relevant.
This is because for $N>0$, excitation manifolds appear where the splitting cannot be observed.
The system can escape in these manifolds using thermal fluctuations. Consequently, the probability $p$ drops below 1, see Fig.~\ref{FIG5qubitcoupling}(b).
We have estimated that the temperature at different bias points varies between 130~mK and 175~mK~\cite{lepp2018Fano}, and is maximal for $N=1$.
Furthermore, the broadening~$\kappa$ increases at $N=1$, since an on-resonance qubit shares photons with the cavity, and thereby can also dissipate them (here with a decay rate much higher than $\gamma$).
This effect is not additive for $N>1$, since at collective resonance it is the average decay rate of all the qubits brought on-resonance, which defines the dissipation rate.
As a result, the signal strength decays more slowly for $N>1$ than between $N=0$ and $N=1$.
However, when increasing $N>1$, another effect comes into play, that instead tends to reduce the broadening~$\kappa$:
the number of qubits that cause broadening of the effective cavity frequency through off-resonance hopping (by inducing fluctuations to the dispersive shift of the cavity~\cite{lepp2018Fano}) reduces.
However, the signal peak becomes narrower and more difficult to detect it from the overall noise.
This behavior of the signal can also be reproduced by master-equation simulations for cavity coupled to eight qubits~\cite{lepp2018Fano}.

The actual limit of maximal observable ensemble size depends on the qubit parameters, chosen bias points and probing tone strength.
Using the calibration scheme, the signal vanishes for six qubits simultaneously on-resonance, and in another cooldown using no calibration and different probing power,
the signal disappears for seven qubits.
Thermal leakage out of the Tavis-Cummings subspace, decay of superconducting qubits, and variations between cooldowns need to be suppressed for coherent control of larger $N$.

\section{\label{sec:conclusion}{Conclusions}}
We have demonstrated the enhancement of the collective coupling between a harmonic oscillator and locally tunable two-level systems. The set of up to six collectively coupled qubits is one of the largest ensemble and the one with largest collective coupling demonstrating the Tavis-Cummings splitting in circuit QED to the best of our knowledge. After the submission we became aware of a related work showing up to $N=10$, but approximately $5$ times less ensemble coupling strength \cite{WangPRL} than our work. The system was realized by a superconducting coplanar resonator coupled to eight frequency-controllable transmons. Our experiment showed that this moderately scaled circuit can be well controlled even in the presence of parasitic effects like background transmission, dissipation, flux control crosstalk, low anharmonicity and elevated sample temperatures, all of which are likely subtle features of near-term physical quantum simulators. A method was presented to calibrate for the crosstalk between the qubits and non-neighbouring flux coils {using a single, shared readout resonator, allowing for precise individual qubit control. The spectroscopic measurement on the collective interaction confirmed that in this system the collective coupling strength scales with $\sqrt{N}$.\\
Increasing the collective coupling opens up the path for further research~\cite{Anton2018}
such as ultra-strong coupling between two modes, ground-state squeezing, and superradiant emission.

\section{\label{sec:acknowledgment}{Acknowledgement}}
The authors thank Lucas Radtke for experimental, and the China Scholarship Council (CSC), Studienstiftung des deutschen Volkes, the European Research Council (ERC-648011), the DFG-Center for Functional Nanostructures National Service Laboratory, Helmholtz IVF grant 'Scalable solid state quantum computing', DFG project INST 121384/138-1 FUGG, Russian Science Foundation (16-12-00095), and Ministry of Education and Science of Russian Federation (K2-2017-081) for grant support.

\appendix
\section{\label{sec:appendixSampleWiring}{Wiring}}
The chip is mounted in an aluminium sample box, and wire bonded to input and output microwave lines. The constant (DC) currents for flux bias control of the qubits are provided via bonds to a printed circuit board. The sample is located inside a cryoperm magnetic shield and cooled down by a dilution refrigerator to around $20\,$mK. A microwave generator is employed when multi-photon transitions are probed dispersively. The signal coming out from the sample goes through two circulators and is amplified by a high-electron-mobility transistor (HEMT) at $4.2\,$K, and further at room temperature, before being measured by the VNA. Every qubit has an individual local flux bias control unit which consists of a DC current source and high-frequency filters at room temperature, and factor 10 current dividers at the $4.2\,$K plate to reduce the overall noise to the quantum chip.\\

\section{\label{sec:appendixSample}{Sample fabrication}}

The sample is patterned in a single step by electron-beam lithography, followed by double angle aluminium deposition (total $80\,$nm) on the intrinsic silicon substrate. The size of the Josephson junction is $100\times100\,$nm$^{2}$ with a critical current of $40.6\,$nA. The oxide barrier is formed by a partial oxygen pressure of $0.0177\,$mbar for 25 minutes for dynamic oxidation. Pictures of the sample is shown in Fig.1(a).

\section{\label{sec:apendixCalibration}{Calibration}}

For an arbitrary selection of two flux lines, the signal traces as shown in Fig.1(b) are not always orthogonal to each other due to finite crosstalk. The slopes of the traces correspond to the mutual inductance matrix elements normalized to the self inductance of the corresponding flux lines and qubits. $M_{xy}$ is the mutual inductance between the $x$ qubit and the $y$ flux bias line. This is a consequence from Eq.2 in the main text. We extract the slopes from linear fits to the data traces. To obtain the full mutual inductance matrix we repeat this measurement scheme 28 times for all combinations of flux lines.

\begin{equation}\label{mutual inductance matrix}
  \begin{pmatrix}
    1 & \frac{M_{1b}}{M_{1a}} & \frac{M_{1c}}{M_{1a}} & \frac{M_{1d}}{M_{1a}} & \frac{M_{1e}}{M_{1a}} & \frac{M_{1f}}{M_{1a}} & \frac{M_{1g}}{M_{1a}} & \frac{M_{1h}}{M_{1a}} \\
    \frac{M_{2a}}{M_{2b}} & 1 & \frac{M_{2c}}{M_{2b}} & \frac{M_{2d}}{M_{2b}} & \frac{M_{2e}}{M_{2b}} & \frac{M_{2f}}{M_{2b}} & \frac{M_{2g}}{M_{2b}} & \frac{M_{2h}}{M_{2b}} \\
    \frac{M_{3a}}{M_{3c}} & \frac{M_{3b}}{M_{3c}} & 1 & \frac{M_{3d}}{M_{3c}} & \frac{M_{3e}}{M_{3c}} & \frac{M_{3f}}{M_{3c}} & \frac{M_{3g}}{M_{3c}} & \frac{M_{3h}}{M_{3c}} \\
    \frac{M_{4a}}{M_{4d}} & \frac{M_{4b}}{M_{4d}} & \frac{M_{4c}}{M_{4d}} & 1 & \frac{M_{4e}}{M_{4d}} & \frac{M_{4f}}{M_{4d}} & \frac{M_{4g}}{M_{4d}} & \frac{M_{4h}}{M_{4d}} \\
    \frac{M_{5a}}{M_{5e}} & \frac{M_{5b}}{M_{5e}} & \frac{M_{5c}}{M_{5e}} & \frac{M_{5d}}{M_{5e}} & 1 & \frac{M_{5f}}{M_{5e}} & \frac{M_{5g}}{M_{5e}} & \frac{M_{5h}}{M_{5e}} \\
    \frac{M_{6a}}{M_{6f}} & \frac{M_{6b}}{M_{6f}} & \frac{M_{6c}}{M_{6f}} & \frac{M_{6d}}{M_{6f}} & \frac{M_{6e}}{M_{6f}} & 1 & \frac{M_{6g}}{M_{6f}} & \frac{M_{6h}}{M_{6f}} \\
    \frac{M_{7a}}{M_{7g}} & \frac{M_{7b}}{M_{7g}} & \frac{M_{7c}}{M_{7g}} & \frac{M_{7d}}{M_{7g}} & \frac{M_{7e}}{M_{7g}} & \frac{M_{7f}}{M_{7g}} & 1 & \frac{M_{7h}}{M_{7g}} \\
    \frac{M_{8a}}{M_{8h}} & \frac{M_{8b}}{M_{8h}} & \frac{M_{8c}}{M_{8h}} & \frac{M_{8d}}{M_{8h}} & \frac{M_{8e}}{M_{8h}} & \frac{M_{8f}}{M_{8h}} & \frac{M_{8g}}{M_{8h}} & 1 \\
  \end{pmatrix}.
\end{equation}

The compensation scheme of the crosstalk is based on counter-currents which are applied to all other qubit coils, while only one qubit is effectively tuned. The counter-currents cancel out the flux in the non-tuned qubits, which therefore stay at a fixed frequency. To obtain the necessary compensation currents a 7 variable linear equation set has to be solved. For example, to tune qubit~1, this function set needs to be solved:

\begin{widetext}
\begin{equation}\label{mutual inductance function}
  \left\{
  \begin{aligned}
    \frac{M_{2a}}{M_{2b}}\Delta I_{a}+
    &\Delta I_{b}+\frac{M_{2c}}{M_{2b}}\Delta I_{c}+\frac{M_{2d}}{M_{2b}}\Delta I_{d}+\frac{M_{2e}}{M_{2b}}\Delta I_{e}+\frac{M_{2f}}{M_{2b}}\Delta I_{f}+\frac{M_{2g}}{M_{2b}}\Delta I_{g}+\frac{M_{2h}}{M_{2b}}\Delta I_{h}=0  \\
    \frac{M_{3a}}{M_{3c}}\Delta I_{a}+
    &\frac{M_{3b}}{M_{3c}}\Delta I_{b}+\Delta I_{c}+\frac{M_{3d}}{M_{3c}}\Delta I_{d}+\frac{M_{3e}}{M_{3c}}\Delta I_{e}+\frac{M_{3f}}{M_{3c}}\Delta I_{f}+\frac{M_{3g}}{M_{3c}}\Delta I_{g}+\frac{M_{3h}}{M_{3c}}\Delta I_{h}=0 \\
    \frac{M_{4a}}{M_{4d}}\Delta I_{a}+
    &\frac{M_{4b}}{M_{4d}}\Delta I_{b}+\frac{M_{4c}}{M_{4d}}\Delta I_{c}+\Delta I_{d}+\frac{M_{4e}}{M_{4d}}\Delta I_{e}+\frac{M_{4f}}{M_{4d}}\Delta I_{f}+\frac{M_{4g}}{M_{4d}}\Delta I_{g}+\frac{M_{4h}}{M_{4d}}\Delta I_{h}=0 \\
    \frac{M_{5a}}{M_{5e}}\Delta I_{a}+
    &\frac{M_{5b}}{M_{5e}}\Delta I_{b}+\frac{M_{5c}}{M_{5e}}\Delta I_{c}+\frac{M_{5d}}{M_{5e}}\Delta I_{d}+\Delta I_{e}+\frac{M_{5f}}{M_{5e}}\Delta I_{f}+\frac{M_{5g}}{M_{5e}}\Delta I_{g}+\frac{M_{5h}}{M_{5e}}\Delta I_{h}=0 \\
    \frac{M_{6a}}{M_{6f}}\Delta I_{a}+
    &\frac{M_{6b}}{M_{6f}}\Delta I_{b}+\frac{M_{6c}}{M_{6f}}\Delta I_{c}+\frac{M_{6d}}{M_{6f}}\Delta I_{d}+\frac{M_{6e}}{M_{6f}}\Delta I_{e}+\Delta I_{f}+\frac{M_{6g}}{M_{6f}}\Delta I_{g}+\frac{M_{6h}}{M_{6f}}\Delta I_{h}=0 \\
    \frac{M_{7a}}{M_{7g}}\Delta I_{a}+
    &\frac{M_{7b}}{M_{7g}}\Delta I_{b}+\frac{M_{7c}}{M_{7g}}\Delta I_{c}+\frac{M_{7d}}{M_{7g}}\Delta I_{d}+\frac{M_{7e}}{M_{7g}}\Delta I_{e}+\frac{M_{7f}}{M_{7g}}\Delta I_{f}+\Delta I_{g}+\frac{M_{7h}}{M_{7g}}\Delta I_{h}=0 \\
    \frac{M_{8a}}{M_{8h}}\Delta I_{a}+
    &\frac{M_{8b}}{M_{8h}}\Delta I_{b}+\frac{M_{8c}}{M_{8h}}\Delta I_{c}+\frac{M_{8d}}{M_{8h}}\Delta I_{d}+\frac{M_{8e}}{M_{8h}}\Delta I_{e}+\frac{M_{8f}}{M_{8h}}\Delta I_{f}+\frac{M_{8g}}{M_{8h}}\Delta I_{g}+\Delta I_{h}=0 \\
  \end{aligned}
  \right.,
\end{equation}
\end{widetext}
where $\Delta I_{b}$,  $\Delta I_{c}$, $\cdots$,  $\Delta I_{g}$,  $\Delta I_{h}$ are the 7 variables. To solve the equation set the relation between these variables and $\Delta I_{a}$ has to be computed to apply the compensation currents for $I_{a}$. In other words, with matrix ~\ref{mutual inductance matrix}, we are able to calibrate out the cross-talk between all the coils. The variation in current $\Delta I$ is used, rather than the absolute value of current $I$. Fig.2(d) shows the result after calibration of (c).  Almost no-tilt indicates there is no residual cross-talk between these two flux bias lines.

\section{subtraction of background from transmission data}\label{App:Fano}

Boundary conditions between the cavity and transmission lines in the presence of a background transmission are~\cite{lepp2018Fano}
\begin{align}
\hat a_{\rm out}(t) &=  \sqrt{\kappa_c} \hat a(t)- \frac{1}{1+2i\epsilon}\hat a_{\rm in}(t)-\frac{2i\epsilon}{1+2i\epsilon}\hat b_{\rm in}(t)\label{eq:boundary1} \\
\hat b_{\rm out}(t) &=  \sqrt{\kappa_c} \hat a(t)- \frac{1}{1+2i\epsilon}\hat b_{\rm in}(t)-\frac{2i\epsilon}{1+2i\epsilon}\hat a_{\rm in}(t)\label{eq:boundary2} \,.
\end{align}
Here operators $\hat a_{\rm in/out}$ describe propagating modes on one side of the two-sided cavity and $\hat b_{\rm in/out}$ on the other side. The cavity mode is described by the operator $\hat a$ and the background coupling by parameter $\epsilon$. We consider here a weak background coupling, i.e. $ \vert \epsilon \vert \ll 1 $.
Assuming that we have measured the output $\langle \hat b_{\rm out}(t) \rangle$, and there is no input from side $b$, and we know $\epsilon$, then the cavity field can be deduced from Eq.~(\ref{eq:boundary2}),
\begin{align}
 \sqrt{\kappa_c} \langle \hat a(t) \rangle &= \langle \hat b_{\rm out}(t) \rangle+\frac{2i\epsilon}{1+2i\epsilon} \langle \hat a_{\rm in}(t)\rangle  \label{eq:boundary3} \,.
\end{align}
Here $\kappa_c$ is the effective coupling between the dressed resonator and the transmission line. 
Since the cavity equation of motion depends only weakly on $\epsilon$~\cite{lepp2018Fano}, it follows that this solution is (up to a constant front factor) also the solution for an output without the presence of a background. The data before and after background removing is shown in Fig.~\ref{FIG5background}. For the original data before background removing, a transformation between Fano-shaped peaks and dips is observed. Fig.~\ref{FIG5background} b)(i.e. Fig.2 b) show the result after background extraction, in which Fano resonances do not appear.

\begin{figure} 
\includegraphics[width=1.0\columnwidth]{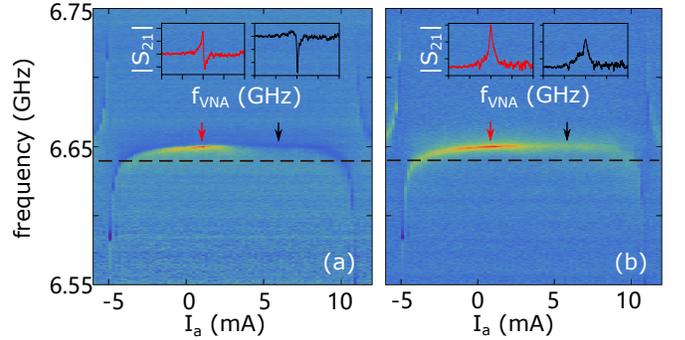}
\caption{Single-flux scan. (a) Original data without background substraction. The two insets show the shape of the resonator when the flux is $1\,$mA and $6\,$mA. (b) The data after background substraction (i.e. Fig.2 b). The two insets show the shape of the resonator when the flux is $1\,$mA and $6\,$mA.}\label{FIG5background}
\end{figure}

\section{\label{sec:appendixMultiple}{Multi-photon transitions}}

The multi-photon transition~\cite{braumuller2015multiphoton} of qubit 7 of the 8-qubit chip in the power spectrum is shown in Fig.~\ref{FIG6anharmonicity}.  All of the qubits are tuned to their maximum frequencies. The VNA is set to the single-photon power and observes the dispersive shift of the resonator while driving qubit 7 separately by a microwave generator. With low driving power, only the fundamental transition is visible. The multi-photon transitions from ground state to higher levels are visible while increasing the power. We determine $\omega^{1,0}_{max}/2\pi=10.24\pm0.08\,$GHz and an anharmonicity $410\pm7\,$MHz (the calculated $E_{c}/2\pi\hbar=462\,$MHz). For transmon qubit, $E_{c}/h=e^{2}/2hC_{total}$, and is approximately the anharmonicity~\cite{koch2007charge}. $E_{J_{max}}/h=34.6\pm0.5\,$GHz is obtained by the maximum frequency of the qubit.

\begin{figure} 
\includegraphics[width=1.0\columnwidth]{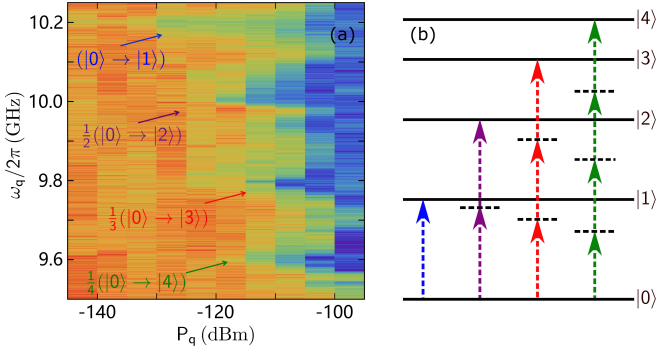}
\caption{Multi-photon transition experiment. (a) Measured qubit frequencies of qubit 7 with increased drive power. At low power, only the fundamental single photon transition from ground state to first excited state is visible. While increasing power, multi-photon transition are observable, and the higher the power, the more transitions show up. The transmitted amplitude is in log-scale. (b) Illustration for the multi-photon transition among the eigen-energy levels of the qubit. }\label{FIG6anharmonicity}
\end{figure}

\section{Extended Jaynes-Cummings model}
In order to explain all features visible in our measurements with one qubit on resonance, we extend the Jaynes-Cummings Model to the case where an anharmonic three level atom is interacting with a bosonic resonator mode. Hamiltonian $H_{3L}$ in Eq.3 in the main text has a block diagonal from, and each block is associated with a fixed conserved number of total excitations in the system consisted by the resonator and 1 qubit. When the total excitation is 0, $H_{3L}^{0}=0$, with basis vector $|g,0\rangle$. When the total excitation is 1,
\begin{equation}\label{H3L excitation=1}
 \hat{H}_{3L}^{1}=
  \begin{pmatrix}
    \omega_{r} & g_{ge}  \\
    g_{ge} & \omega_{e}  \\
  \end{pmatrix},
\end{equation}
with basis vectors $\{|g,1\rangle,\ |e,0\rangle\}$. And when the total excitation is 2,
\begin{equation}\label{H3L excitation=2}
 \hat{H}_{3L}^{2}=
  \begin{pmatrix}
    2\omega_{r} & \sqrt{2}g_{ge} &  0 \\
   \sqrt{2}g_{ge} & \omega_{r}+\omega_{e} &g_{ef}  \\
   0 & g_{ef} & \omega_{f} \\
  \end{pmatrix},
\end{equation}
with basis vectors $\{|g,2\rangle,\ |e,1\rangle,\ |f,0\rangle\}$.
Diagonalization of the Hamiltonians in Eq.~\ref{H3L excitation=1} and Eq.~\ref{H3L excitation=2} yields the eigenenergies of the first two excitation manifolds of the system that is indicated in Fig.3 c.

\section{Fitting the splitting}\label{App fitting}
Consider a single two-level qubit couples to a resonator, the Hamiltonian is the same as Eq.~\ref{H3L excitation=1}. The eigenvalues of this Hamiltonian are
\begin{equation}\label{eigenvalues JC}
 \frac{E_{\pm}}{\hbar}=\frac{\omega_{r}+\omega_{e}}{2}\pm\frac{1}{2}\sqrt{4g_{ge}^{2}+(\omega_{r}-\omega_{e})^{2}}
\end{equation}
In the vicinity of the strong coupling to the resonator range, the relation between qubit energy and the applied flux bias current is simplified to a linear function $\omega_{e}(I)=2\pi(aI+b)$. By substitution into Eq.~\ref{eigenvalues JC}, one gets the fitting function for a single qubit interacting with the resonator.
\begin{equation}\label{eigen frequency JC}
 f_{\pm}(I)=\frac{f_{r}+aI+b}{2}\pm\frac{1}{2}\sqrt{4(\frac{g_{ge}}{2\pi})^{2}+(f_{r}-aI-b)^{2}}.
\end{equation}
For multiple qubit case, treating them as an ensemble (ens), the effective total coupling strength is enhanced. In order to obtain the value, the multiple-qubit anticrossing is fitted with the following formula:
\begin{equation}
\centering
  \begin{aligned}
f(I)_{ens_{+}}=&\frac{f_{r}+aI+b}{2}+\frac{1}{2}\sqrt{4(\frac{g_{ge}}{2\pi})^{2}+(f_{r}-aI-b)^{2}},   \\
f(I)_{ens_{-}}=&\frac{f_{r}+a(I+I_{shift})+b}{2}-f_{shift} \\
        &-\frac{1}{2}\sqrt{4(\frac{g_{ge}}{2\pi})^{2}+[f_{r}-a(I+I_{shift})-b]^{2}}. \\
  \end{aligned}
\label{Equation fit function}
\end{equation}
Eq.~\ref{Equation fit function} has the same form as Eq.~\ref{eigen frequency JC} but the lower branch of the anticrossing has two more degrees of freedom ($I_{shift}$, $f_{shift}$) which shift its position compared to the single qubit anticrossing. The effective coupling strength is extracted by the minimum distance between these two branches (i.e. the ensemble and the resonator are exactly on-resonance).
\begin{equation}\label{effective coupling strength}
 \frac{g_{ens}(I)}{2\pi}=\frac{f(\frac{f_{r}-b}{a}-\frac{I_{shift}}{2})_{ens_{+}}-f(\frac{f_{r}-b}{a}-\frac{I_{shift}}{2})_{ens_{-}}}{2}
\end{equation}
The result of fitting the data to Eq.~\ref{effective coupling strength} is plotted by red dots in Fig.4 (a).

\section{Analysis of signal strength}
Assuming that the system is with a probability $p$ in a state which allows for observing the studied transition,
and that there is an internal broadening of the dressed system $\gamma_{eff}$, the average transmission around the
corresponding resonance frequency $\omega_0$  has the form~\cite{lepp2018Fano}
\begin{eqnarray}
s_{12}(\omega)=\frac{2\epsilon}{{\rm i} - 2\epsilon} + p\frac{\kappa_c}{\kappa_c+\gamma_{eff} + {\rm i}(\omega_0 - \omega)} \,.
\end{eqnarray}
Here $\kappa_c$ is the effective coupling between the dressed resonator and the transmission line. 
Since we can measure $\epsilon$ from the off-resonance transmission, this formula can be used to extract parameters $p$ and $\gamma_{eff}$ from the experimental data.  
We also plot the effective signal strength $p\frac{\kappa_c}{\kappa_c+\gamma_{eff}}$ in Fig.4 e).



\end{document}